\documentclass[twocolumn,showpacs,amsmath,amssymb,pra,nofootinbib]{revtex4}

\usepackage{graphicx}
\usepackage{dcolumn}
\usepackage{bm}
\usepackage{times,mathptm}
\usepackage[ps2pdf,colorlinks]{hyperref}

\newcommand{\ket}[1]{|#1\rangle}
\newcommand{\bra}[1]{\langle#1|}
\newcommand{\script}[1]{{\cal{#1}}}

    {
    \smallskip
    \refstepcounter{theorem}
    \noindent
    {\bf Example \arabic{section}.\arabic{theorem}} \ \ }
    {\hspace*{\fill}{$\Diamond$}
    \smallskip}

    {
    \smallskip
    \refstepcounter{theorem}
    \noindent
    {\bf Remark \arabic{section}.\arabic{theorem}} \ \ }
    {\hspace*{\fill}{$\Diamond$}
    \smallskip}

    {
    \smallskip
    \refstepcounter{theorem}
    \noindent
    {\bf Definition \arabic{section}.\arabic{theorem}} \ \ }
    {\hspace*{\fill}{\ }
    \smallskip}

    {
    \smallskip
    \refstepcounter{theorem}
    \noindent
    {\bf Definition \Alph{section}.\arabic{theorem}} \ \ }
    {\hspace*{\fill}{\ }
    \smallskip}

    {
    \smallskip
    \refstepcounter{theorem}
    \noindent
    {\bf Scholium \arabic{section}.\arabic{theorem}} \it \ \ }
    {\hspace*{\fill}{\ }
    \smallskip}

\newenvironment{proof}[1][]
    {
    \noindent
    {\bf Proof{#1}:  }
    }
    {\hspace*{\fill}{$\Box$}\smallskip}

    {
    \noindent
    {\bf Sketch{#1}:  }
    }
    {\hspace*{\fill}{$\Box$}\smallskip}

    {
    \noindent
    {\bf Reference:  }
    }
    {\hspace*{\fill}{$\odot$}\smallskip}

\newtheorem{theorem}{Theorem}[section]

\newtheorem{lemma}[theorem]{Lemma}

\begin{document}
\bibliographystyle{apsrev}

\title{Criteria for Exact Qudit Universality}

\author{Gavin K. Brennen$^1$}\email{gavin.brennen@nist.gov}
\author{ Dianne P. O'Leary$^{2,3}$}\email{oleary@cs.umd.edu}
\author{Stephen S. Bullock$^3$} \email{stephen.bullock@nist.gov}

\affiliation{
$^1$ National Institute of Standards and Technology, Atomic Physics Division,
Gaithersburg, Maryland, 20899-8420 \\
$^2$ University of Maryland, Department of Computer Science, Collge Park,
Maryland 20742 \\
$^3$ National Institute of Standards and Technology, Mathematical and 
Computational Sciences Division, Gaithersburg, Maryland 20899-8910
}
\date{\today}

\setlength{\unitlength}{0.125in} 

\begin{abstract}
We describe criteria for implementation of quantum computation in qudits.  A qudit is a $d$-dimensional system whose Hilbert space is spanned by states $\ket{0}$,
$\ket{1}$, $\ldots$, $\ket{d-1}$.  An important earlier work of Mathukrishnan
and Stroud \cite{MathukrishnanStroud:00} describes how to exactly simulate an arbitrary unitary on multiple 
qudits using a $2d-1$ parameter family of single qudit and two qudit gates.  Their technique 
is based on the spectral decomposition of unitaries.  Here we generalize this argument
to show that exact universality follows given a discrete set of single qudit Hamiltonians and one 
two-qudit Hamiltonian.  The technique is related to the $QR$-matrix decomposition
of numerical linear algebra.  We consider
a generic physical system in which the single qudit Hamiltonians are
a small collection of $H_{jk}^x=\hbar\Omega (\ket{k}\bra{j}+\ket{j}\bra{k})$
and $H_{jk}^y =\hbar\Omega (i\ket{k}\bra{j}-i\ket{j}\bra{k})$.  
A coupling graph results taking nodes $0$, $\ldots$, $d-1$ and edges
$j \leftrightarrow k$ iff $H_{jk}^{x,y}$ are allowed Hamiltonians.  One qudit
exact universality follows iff this graph is connected, and complete
universality results if the two-qudit Hamiltonian
$H=-\hbar\Omega \ket{d-1,d-1}\bra{d-1,d-1}$ is also allowed.  We discuss implementation in the eight dimensional ground electronic states of $^{87}$Rb and construct an optimal gate sequence using Raman laser pulses.
\end{abstract}



\maketitle
\section{Introduction}

\noindent
An important theoretic construct used in the field of quantum information is the qubit.  Its utility follows from the simple but significant recognition that all two dimensional subspaces, regardless of the underlying physical system, can be regarded as informationally equivalent.  This has made it possible to discuss quantum computation in terms of single qubit and two qubit gates without the need to analyze the specific interactions that realize operations within a physical system or between subsystems.  An important issue in this regard is that a necessary condition for {\em efficient} quantum computation is the existence of an underlying tensor product structure on the Hilbert space $\mathcal{H}$.  If all computation were performed on a single $d=dim(\mathcal{H})$ level system then some physical resource such as space or energy would grow with the dimension of the system \cite{Blume-KahoutEtAl:02}.  In contrast, the analogous resources grow poly-logarithmically with the dimension when the system is composed of many subsystems.  By this argument, a computation performed on qubits $(d=2)$ is in some sense the most efficient foliation of Hilbert space.  

Nevertheless, there are compelling reasons to consider computation on qudits with $d>2$.  First,
most physical implementations encode qubits in a subspace of a larger Hilbert space.  Using higher dimensional subspaces already endowed in these systems may be more efficient in terms of the number of interacting gates needed for an algorithm that acts on a Hilbert space of fixed dimension.  This is critical for error control because interactions between qudits tend to open channels for interactions with the decohering environment.  By contrast, in many physical systems, single qudit control is a well developed technology that can be done with high precision.  Second, there is some evidence that the error thresholds for fault tolerant computation improve when the encoding is done with qudits where $d>2$ and prime \cite{Aharonov}.

Previous work by Brylinski and Brylinski proves the necessary and sufficient criteria for exact qudit universality \cite{Brylinski:02}.   Exact universality means that any unitary and, by unitary extension to a larger Hilbert space, any {\em quantum process}, can be simulated with zero error.  The authors show that arbitrary single qudit gates complemented by one entangling two qudit gate is needed.  Their method is not constructive.  Muthukrishnan and Stroud \cite{MathukrishnanStroud:00} give a constructive procedure for an exact simulation of an arbitrary unitary on $n$ qudits using single qudit and two qudit gates.  Their approach uses the spectral decomposition of unitaries and involves a gate library consisting of a family of continuous parameter gates.  Here we describe an approach that uses the $QR$ decompositions on unitaries to achieve exact universal computation on qudits.  This construction has the advantage that  the single qudit gates are generated by a fixed set of Hamiltonians that couple pairs of states in the single qudit logical basis.  The gates perform rotations, parameterized by one angle, about orthogonal axes within the associated two dimensional subspace.  Additionally, our decomposition requires only one fixed two qudit gate, the controlled increment gate $(\mathtt{CINC})$ gate.  This gate can be simulated by at most $d-1$ instances of a two-qudit Hamiltonian $H_{int}$ that generates a phase on a single product state of two qudits.  Such interactions can be engineered in many atom optical systems.  

In this paper, the general results are developed with close contact to the example of computation in the $d=8$ qudit encoded in the ground hyperfine states of $^{87}$Rb.  In Sec. \ref{sec:1qudit} we describe the construction of single qudit unitaries using the $QR$ decomposition.  We introduce a coupling graph to describe how states are connected to each other by physical Hamiltonians.  The set of rotation planes may be incomplete, i.e. each state may not be connected to every other state.  However, provided the graph is connected, an efficient decomposition can be found.  Multiqudit computation is addressed in Sec. \ref{sec:multiqudit}.  It is shown that a single two-qudit gate when combined with single qudit gates suffices to generate arbitrary two qudit unitaries and hence completes the requirements for exact universality.  In the appendix we show how to convert between the gate library introduced here and the family of gates used in Ref. \cite{MathukrishnanStroud:00}.  Finally, we conclude with a summary of the results in Sec. \ref{sec:conc}.

\section{One-qudit unitaries}
\label{sec:1qudit}

We pick a fixed gate library for single qudit operations involving rotations about non orthogonal axes of two dimensional subspaces.  Within each subspace $\mathcal{H}_{jk}$, the gates are generated by the two Hamiltonians:
\begin{equation}
H_{jk}^x=\hbar\Omega(\ket{j}\bra{k}+\ket{k}\bra{j}),\quad
H_{jk}^y=\hbar\Omega(-i\ket{j}\bra{k}+i\ket{k}\bra{j}).
\label{singlecouple}
\end{equation}
For convenience of notation, we assume the strength of each coupling is equal to $\Omega$ and leave the time each Hamiltonian is applied as a free parameter.  Any unitary in the two dimensional subspace can be written
\begin{equation}
\begin{array}{llll}
U_{j,k}(\gamma,\phi,\theta)
& \equiv &
\exp \bigg[ \; & -i\gamma (\sin(\theta)\cos(\phi)
\big(\; \ket{j}\bra{k}+\ket{k}\bra{j}\; \big) \\
& & & +\; \sin(\theta)\sin(\phi)\big( 
-i\ket{j}\bra{k}+i\ket{k}\bra{j}\; \big)\\
& & & +\; \cos(\theta)\big( \; \ket{j}\bra{j}-\ket{k}\bra{k}\; \big)
\; \bigg] \\
&=& & e^{-iH^x_{jk}t/\hbar}
\; e^{-iH^y_{jk}t^{\prime}/\hbar}\; e^{-iH^x_{jk}t^{\prime\prime}/\hbar}
\label{evolve}
\end{array}
\end{equation}
for the appropriate $t,t^{\prime},t^{\prime\prime}$ using the XYX Euler angle decomposition \cite{NielsenChuang:00}.
In some cases, the two Hamiltonians in Eq. \ref{singlecouple} can be turned on simultaneously.  By adjusting the relative strengths of the couplings, one can then realize any rotation about an axis on the equator of the Bloch sphere in the two dimensional subspace.  For brevity, we write $U_{jk}(\gamma,\phi)=e^{-i(\cos(\phi)H^x_{jk}+\sin(\phi)H^y_{jk})\gamma/(\hbar\Omega)}$, where it is understood that if the couplings $H^x_{jk},H^y_{jk}$ cannot be turned on together then $U_{jk}(\gamma,\phi)$ requires three elementary gates.

Realization of an arbitrary unitary evolution $v \in U(d^n)$ requires
two steps.  The first corresponds to a $QR$ decomposition \cite{Cybenko:01}
of the matrix $v$.
\begin{itemize}
\item  Using the allowed set of Hamiltonians, we may realize matrices
of \emph{Givens rotations} physically.  These are described in the
next paragraph.  Generically, the $QR$ decomposition writes an invertible
$G=UT$, where $U=G_1G_2 \ldots G_\ell$ is a product of Givens rotations and
hence unitary and $T$ is upper triangular.  Note that if $G=V$ is unitary,
then so likewise is $T=U^\dagger G$, whence $T$ is in this case a diagonal
matrix which applies relative phases to computational basis states.
\item  Using techniques for realizing diagonal computations 
\cite{BullockMarkov:04}, a sequence of 
Hamiltonians realizing $T$ is constructed.
\end{itemize}

We next illustrate the idea of a Givens rotation by way of example,
retaining $V$ as above.  We may choose $U$ a Givens rotation so as
to zero the matrix element $(UV)_{d-1,0}$ (where the indices run $0,1,\cdots,d-1$.)
Specifically, suppose that
\begin{equation}
\label{eq:givens}
U_{d-1,d-2}(\gamma,\phi) = 
\left(
\begin{array}{rrrr}
I_{d^n-2} & &  & \\ & \ddots &  &  \\
 & & \cos({\gamma}) & -ie^{i\phi}\sin({\gamma}) \\
 & &  -ie^{-i\phi}\sin({\gamma}) & \cos({\gamma}) \\
\end{array}
\right)
\end{equation}
Here, we choose the angles $\gamma$, $\phi$ as follows:
\begin{equation}
\begin{array}{lcl}
\tan{\gamma}&=&
|v_{d-1,0}/v_{d-2,0}| \\
\phi&=&\pi/2+\arg(v_{d-2,0})-\arg(v_{d-1,0}) \\
\end{array}
\end{equation}
where $v_{m,n}$ are the entrees of the unitary $V$.  Then letting $v'_{\ast \ast}$ denote a changed entry, we obtain:
\begin{equation}
U_{d-1,d-2} V = 
\left(
\begin{array}{rrrr}
v_{0,0} & v_{0,1} & \cdots & v_{0,d-1} \\
\vdots & \ddots & \cdots & \vdots \\
v^{\prime}_{d-2,0} & \cdots & v^{\prime}_{d-2,d-2} & v^{\prime}_{d-2,d-1} \\
0 & \cdots & v^{\prime}_{d-1,d-2} & v^{\prime}_{d-1,d-1} \\
\end{array}
\right)
\end{equation}
In the next step, one chooses a unitary $U_{d-1,d-3}$ to zero the matrix element $(U_{d-1,d-2}V)_{d-2,0}$.  Continuing carefully in this way allows one to complete the $QR$
decomposition described above by introducing a zero into every entry of the 
resulting unitary below the diagonal.

\subsection{Example: One-qudit Unitaries in $^{87}${Rb}}
\label{sec:ex}

We begin by describing explicitly the implications of our constructions
for an example which is related to but not covered explicitly by earlier
work of Mathukrishnan and Stroud \cite{MathukrishnanStroud:00}.
Specifically, we describe the coupling graph alluded to in the
introduction in this case before defining it in general.

\begin{figure}
\begin{center}
\includegraphics[scale=0.33]{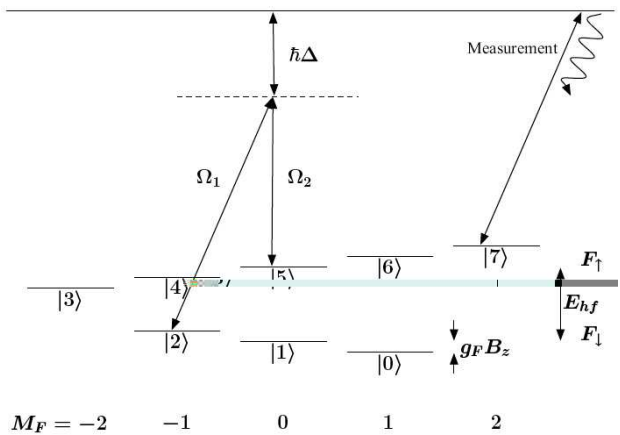}
\caption{\label{fig:hyperfine}A single $d=8$ 
qudit (quoctet) encoded in the ground 
state hyperfine levels of $^{87}$Rb.  A pair of lasers can couple states 
in different hyperfine manifolds according to the selection rule 
$\Delta M_F=0,\pm1$.  Projective measurements of population in state 
$|7\rangle$ are made by observing resonant fluorescence on a cycling 
transition to the excited state.  Any pair of states can be coupled by 
swapping neighbors together pairwise and similarly any state can be measured 
by swapping to $|7\rangle$.}
\end{center}
\end{figure}

Thus, consider the 
atomic species $^{87}$Rb per Fig. \ref{fig:hyperfine}.  
There are two ground state 
hyperfine manifolds with total spin $F_{\downarrow}=1$ and 
$F_{\uparrow}=2$ split in energy by the hyperfine interaction 
$E_{hf}$.  Each manifold consists of $2F+1$ degenerate magnetic sublevels 
$M_F$ for a total of eight distinguishable states.  We designate this 
$d=8$ system a quoctet.  The degeneracy can be lifted by applying a 
longitudinal magnetic field $B_z$.  For small fields, the resultant Zeeman interaction 
is linear in the magnetic quantum number:  $H_B=g_FB_z M_F$, where the Lande 
$g$ factors satisfy $g_{F_{\downarrow}}=-g_{F_{\uparrow}}$ \cite{footnote}.

There are several ways to couple the magnetic sublevels including the use of microwave pulses and 
Raman lasers.  These techniques are usually distinguished by the strength of the coupling with respect to the hyperfine interaction.  We consider coupling that is weak relative to $E_{hf}$ using a pair of laser beams on Raman resonance between two sublevels at a time.
The effective atom-laser Hamiltonian $H_{AL}$ in the subspace $\mathcal{H}_{jk}$ is:
\begin{equation}
\label{eq:atomlaser}
H_{ALjk} =  
\cos(\phi)H^x_{jk}+\sin(\phi)H^y_{jk}
\end{equation}
where $\Omega=|\Omega_1\Omega_2|/\Delta$ is the product of the individual 
laser Rabi frequencies divided by the detuning $\Delta$ from the excited 
state, and $\phi=\phi_1-\phi_2$ is the relative phase of the two beams. 
In order to selectively couple two states it is necessary that their energy difference 
be unique.  In the linear Zeeman regime, this can only be accommodated when the two levels 
reside in different hyperfine manifolds.  Additionally, it will be important to minimize spontaneous 
emission during the pulse sequence by choosing a large detuning $\Delta$ of each laser from the 
excited states.  The allowed couplings are constrained by angular momentum selection rules which dictate the change in magnetic spin quantum number during a single pulse sequence.  For detuning $\Delta$ much 
greater than the excited state hyperfine structure, but less than than the fine structure splitting, the angular momentum selection rules dictate $\Delta M_F=0,\pm 1$ and $\ket{F_{\downarrow},0}\nleftrightarrow \ket{F_{\uparrow},0}$.  Using two-laser 
pulses of the appropriate frequency and polarization, the 
states $|F_{\downarrow},M_F\rangle$ and $|F_{\uparrow},M_F+\Delta M_F\rangle$ 
where $\Delta M_F=0,\pm 1$ can then be coupled together.  This is shown 
schematically in Fig. \ref{fig:hyperfine} where states $|2\rangle$ and 
$|5\rangle$ are coupled by a $\sigma_+-\pi$ polarized laser pair. 

At this point we pause to comment on the resources necessary for single quoctet compution using  Raman pulses.  Transitions realizing $\Delta M_F=0,\pm 1$ can be achieved by choosing the correct polarizations for the lasers with respect to a quantization axis defined by the magnetic field direction.  For a fixed Zeeman splitting, it will be necessary to have lasers tuned to Raman resonance for all the allowed couplings.  This may be achievable using a fixed source laser source that is frequency modulated appropriately.  Another recourse is to change the magnetic field strength for each pairwise state coupling so that only one laser pair of fixed frequency is necessary.  The phase shifts accumulated on the basis states during the change in Zeeman interaction can be accounted for in the gate sequence. 

We wish to show that the above set of atom-laser Hamiltonians suffices to
construct an arbitrary unitary evolution of the one-quoctet phase space
\hbox{$\mathcal{H}_1 = 
\mathbb{C}\ket{0}\oplus \cdots \oplus \mathbb{C}\ket{7}$}.
Take $V \in U(8)$ as the target one-quoctet evolution, where
$U(8)$ is the symmetry group of the inner-product on the Hilbert space
(i.e. $V V^\dagger=I_8$.)  The goal then is to decompose $V$ into a sequence
of evolutions by these atom laser Hamiltonians:
\begin{equation}
V = \mbox{exp}(i H_{AL}^1t_1/\hbar) \cdots \mbox{exp}(i H_{AL}^\ell t_\ell/\hbar)
\end{equation}
Additionally, we prefer \emph{efficient} decompositions, i.e. we wish to
use as few laser pulses (as small an $\ell$) as possible.  This is sometimes
not possible, depending on which states $\ket{j}$, $\ket{k}$ are coupled
by an $H_{AL}$.  In order to classify when the $QR$ step is possible,
we introduce the notion of a \emph{coupling graph}, by example.

\vbox{
\noindent
{\bf $^{87}$Rb coupling graph:}
The $^{87}$Rb coupling graph has vertices labelled by $0,1,\ldots,7$.
In addition, consulting Fig. \ref{fig:hyperfine}, we also allow in the
following edges, corresponding to the atom-laser coupled hyperfine states.
\begin{equation}
\big\{ (0,5), (0,6), (0,7), (1,4), (1,6), (2,3), (2,4), (2,5) \big\}
\end{equation}
In particular, the edges encode the selection rule for the hyperfine states.
The graph is reproduced in Fig. \ref{fig:Hamiltoniangraph}.  We note for
future use that it is connected.  Provided the states $\ket{j}$, $\ket{k}$ are coupled, 
we may produce any determinant-one unitary evolution of $\mathcal{H}_{jk}$ 
using Eq. \ref{evolve}.
}

\setlength{\unitlength}{0.025cm}
\begin{figure}
\begin{picture}(250,160)
\put(100,140){\large \bf $0$}
\put(155,140){\large \bf $7$}
\put( 60, 95){\large \bf $6$}
\put(140, 95){\large \bf $5$}
\put( 60, 50){\large \bf $1$}
\put(140, 50){\large \bf $2$}
\put(195, 50){\large \bf $3$}
\put(100,  5){\large \bf $4$}
\put(110,145){\line(1,0){40}} 
\put(150, 55){\line(1,0){40}}  
\put( 62, 65){\line(0,1){20}}  
\put(142, 65){\line(0,1){20}}  
\put( 95,135){\line(-1,-1){25}} 
\put(110,135){\line( 1,-1){25}} 
\put( 95, 15){\line(-1, 1){25}} 
\put(110, 15){\line( 1, 1){25}} 
\end{picture}
\caption{\label{fig:Hamiltoniangraph} This is the coupling graph for the
coupled hyperfine states of $^{87}$Rb (cf. Fig. \ref{fig:hyperfine}.)
As it is connected, the collection of atom-laser couplings allows for universal
one-quoctet computation.}
\end{figure}

\setlength{\unitlength}{0.125in}

Now note that \emph{since the coupling graph is connected,} we may
in fact sequentially construct a Givens rotation on any $\mathcal{H}_{jk}$.
Indeed, even if $\ket{j}$ and $\ket{k}$ are not paired, there exists
a sequence $\ket{j_0}=\ket{j}, \ket{j_1},\ket{j_2},\ldots,\ket{j_\ell}=\ket{k}$
such that each consecutive pair admits atom-laser Hamiltonians.  Moreover,
taking $\phi=\pi/2, \theta=\pi/2$ in Equation \ref{eq:givens} shows that
we may use these pairings to swap states up to relative phase.
Hence, since we may physically construct some sequence of Hamiltonians for
any Givens rotation, we see that the first step of the $QR$ decomposition
is possible.

This leaves open the question of efficiency.  For example, one might hope
that in a graph as highly connected as that for $^{87}$Rb few or no
swaps might be required.  This is indeed possible as we now show.  It is convenient to reorder the unitary in a logical basis 
labeled $\{7,0,6,5,3,2,4,1\}$.  By 
successive Givens rotations, one may bring a unitary $V$ to diagonal form 
column by column where the
sequence is chosen so as to not void zeroes created in earlier steps.  Each of the columns can be reduced to a single unimodular entry on the diagonal by a sequence of Givens rotations $U_{j,k}$ acting on the two dimensional 
subspace $\mathcal{H}_{jk}$ as follows \cite{O'Leary}:  

\begin{itemize}

\item Column 7 reduction:
$
U_{4,1}U_{2,4}U_{2,3}U_{5,2}U_{0,5}U_{0,6}U_{7,0}
$
\item Column 0 reduction:
$
U_{4,1}U_{2,4}U_{2,3}U_{5,2}U_{0,5}U_{0,6}
$
\item Column 6 reduction:
$
U_{2,5}U_{2,3}U_{4,2}U_{1,4}U_{6,1}
$
\item Column 5 reduction:
$
U_{4,1}U_{2,4}U_{2,3}U_{5,2}
$
\item Column 3 reduction:
$
U_{4,1}U_{2,4}U_{3,2}
$
\item Column 2 reduction:
$
U_{4,1}U_{2,4}
$
\item Column 4 reduction:
$
U_{4,1}
$
\end{itemize}  

Note that in general,
constructing $U_{jk}$ requires $2d(j,k)-1$ basic Hamiltonians, where 
$d(j,k)$ is the distance between $j$ and $k$ in the graph
corresponding to the pairing relation.  For qudit computation in $^{87}$Rb using Raman
pulses, the graph is sufficiently connected so that the distance is never greater than one in the 
QR decomposition above. The are a total of $8\times 7/2=28$ gates in the reduction to diagonal form.  Each gate $U_{j,k}\in SU(2)$ has two parameters so this gives 56 parameters.  An arbitrary $u\in SU(d)$ requires $d^2-1$ parameters
so the additional 7 parameters correspond to seven relative phases 
left on the diagonal.

\subsection{Relative Phases}

The goal of this section is to show that should the Hamiltonian
graph be connected and $T=\sum_{j=0}^{d-1} \mbox{e}^{i \varphi_j}
\ket{j}\bra{j}$ be a diagonal element of $U(d)$, then we may realize
$T$ with the allowed Hamiltonians $H^x_{jk}$, $H^y_{jk}$.  In fact, we only need to construct 
$T$ up to a global phase so we can simulate $T^{\prime}=e^{i\phi_{d-1}}T$.  We first
note that although it is not explicitly an allowed Hamiltonian, we may
for any $(j,k)$-edge within the coupling graph simulate the
effect of $H^z_{jk} =\hbar\Omega (\ket{j}\bra{j}-\ket{k}\bra{k})$.  Indeed,
for any fixed angle $\gamma$ we have
\begin{equation}
e^{-iH^z_{jk}\gamma/(\hbar\Omega)}=U_{j,k}(-\pi/4,\pi/2)U_{j,k}(\gamma,0)U_{j,k}(\pi/4,\pi/2).\;
\end{equation} 
The goal then 
is to find a sequence of $z$ rotations that simulates $T^{\prime}$:
\begin{equation}
\prod_{l=1}^p \exp(-iH^z_{j_lk_l}t_l/\hbar)=T^{\prime}\;
\label{diagsim}
\end{equation}
Given that the coupling graph is connected, choose a subset $S$ of $d-1$ edges $\lambda^z_{jk}=\ket{j}\bra{j}-\ket{k}\bra{k}$
that leave the graph connected.  We can represent the elements of $S$ as vectors in a $d$ dimensional
real vector space spanned by the orthonormal vectors $\{e_j\}$, i.e. $\lambda^z_{jk}=e_j-e_k$.
We then construct a $(d-1)\times d$ matrix $M$ out of the row vectors in $S$:    $M=\{\lambda^z_{0k_0},\lambda^z_{1k_1},\ldots\lambda^z_{d-2k_{d-2}}\}$.  The appropriate timings
$t_j$ in Eq. \ref{diagsim} necessary to simulate $T^{\prime}$ are given by solutions to the matrix equation $M \vec{\theta}=\vec{\phi}$, where $\vec{\theta}=\Omega \{t_0,\ldots t_{d-2}\}$ and $\vec{\phi}=\{\phi_0,\ldots \phi_{d-2}\}$.  The angle $\phi_{d-1}=0$ for the unitary $T^{\prime}$ by assumption.  It is easily verified by Gaussian elimination that the dimension of the row space of $M$ is $d-1$, thus there is a unique solution to the vector $\vec{\theta}$.

The result is that any diagonal unitary can be simulated up to a global phase using $3\times (d-1)$ gates from the gate library.  This sequence can be reduced by a factor of three if $z$ rotations can be implemented directly without conjugation.  Further, all the Hamiltonians $H_{jk}^z$ are diagonal and hence commute, so $z$ rotations that act on disjoint subspaces can be implemented in parallel using additional control resources.      

\subsection{One-qudit universality for generic coupling graphs}
\label{sec:1quditun}


We found that for computation in the $^{87}$Rb quoctet, a single qudit unitary could be 
brought to diagonal form using the fewest possible Givens rotations.  This is not peculiar to that system but is in fact possible for any system with a connected
coupling graph.

\begin{lemma}[\cite{O'Leary}]
\label{lem}
Given a $d$-node coupling graph $\script{G}$ of
allowed Givens rotations,
then any $U \in SU(d)$ can be brought to diagonal form
using $d(d-1)/2$ allowed rotations
if and only if $\script{G}$ is connected.
\end{lemma}

\begin{proof}
Suppose $\script{G}$ is connected. 
Form any spanning tree for it, and renumber the nodes so that
the path from node $d$ (the root of the tree)
to any node $j$ passes through no node numbered lower than 
$j$;  such a numbering can be constructed by successively
deleting leaf nodes and numbering in order of deletion.
(For $^{87}${Rb}, we formed the tree by breaking the edge 
between nodes 6 and 1 and used the logical basis ordering
$\{7,0,6,5,3,2,4,1\}$.)
At the $j$th step ($j=1,\dots,d-1$), 
create the tree $\script{T_j}$, rooted at node $j$, from the portion
of the spanning tree defined by nodes $j,\dots,d$.
(Note that $\script{T_j}$ is connected due to the way we numbered
the nodes.)
Then, until only the root of $\script{T_j}$
remains, choose a leaf $k$, use a rotation defined by its edge to
eliminate element $(k,j)$ of $U$, and delete node $k$
from $\script{T_j}$.
The result of applying these steps is an upper triangular 
matrix (and therefore, since $U$ is unitary, a diagonal
matrix) computed by using $d(d-1)/2$ allowed rotations.

Suppose $\script{G}$ is not connected and consider a matrix
$U \in SU(d)$ that has 
no zero elements.  Choose an arbitrary
node to call node $1$.
Then we can at best eliminate all but one of the nonzeros
in column 1 of the disconnected piece, but there is no
allowed rotation that will eliminate the last nonzero.
Repeating the argument for each choice of node $1$, we 
conclude that we cannot reduce $U$ to diagonal form
using only allowed rotations.
\end{proof}

\section{Multi-qudit Universaility}
\label{sec:multiqudit}

Suppose in addition to being allowed local Hamiltonians 
$\{H_{jk}^{x,y}\}$ with a connected coupling graph, the
physical system also allows for a two-qudit phase Hamiltonian
\begin{equation}
H_{int}=-\hbar\Omega \ket{d-1,d-1} \bra{d-1,d-1}.\;
\label{couple}
\end{equation}
Using known but dispersed techniques
\cite{BarencoEtAl:95,MathukrishnanStroud:00}, we describe a bootstrap
which allows for universal quantum computation.  Note that due to the
standard $QR$ decomposition, it suffices to construct arbitrary two-qudit
unitary evolutions \cite{MathukrishnanStroud:00, Cybenko:01}.  In fact arbitrary one-qudit operations controlled on $d-1$, i.e.
$\Lambda_1(v) = I_{d^2-d}\oplus v \in U(d^2)$ for $v \in U(d)$, suffice.

Before presenting the generic discussion, we describe a particular
example of a two-qubit operation which has seen heavy use 
\cite{MathukrishnanStroud:00}.  First, we label as 
$(\oplus 1)$ the self-map of $\mathbb{Z}/d \mathbb{Z}$ which carries
$k \mapsto (k+1) \mbox{mod }d$.  Then the controlled-increment gate,
abbreviated here as {\tt CINC}, is defined by extending the following
rule linearly:
\begin{equation}
{\tt CINC} \ket{j,k} = \left\{
\begin{array}{rr}
\ket{j,k}, & j \neq d-1 \\
\ket{j,k\oplus 1}, & j = d-1 \\
\end{array}
\right.
.
\end{equation} 
The {\tt CINC} gate is heavily used in the literature in 
building a generic $k$-controlled
computation $\Lambda_k(v)$  \cite{MathukrishnanStroud:00} as well as for
constructing quantum error correction codes \cite{Grassl:03}.

We may explicitly realize {\tt CINC} from the Hamiltonian
$H_{int}= -\Omega \ket{d-1,d-1}\bra{d-1,d-1}$ as follows.
This discussion uses the group theory notation that for $j_1,\ldots,j_\ell
\subset \{0,1,\ldots,d-1\}$,
we write $(j_1j_2\ldots j_\ell)$ for the cyclic permutation with
$j_1 \mapsto j_2$, $j_2 \mapsto j_3$, $\cdots$, $j_{\ell-1}\mapsto j_\ell$,
$j_\ell \mapsto j_1$, and all other set elements fixed.
The permutation will also be identified implicitly with the associated
permutation matrix $\pi_{(j_1 j_2 \ldots j_\ell)} \in U(d)$.  Hence,
given $(01)(12)\cdots(d-2\; d-1) = \oplus 1$, we see that
${\tt CINC} = \Lambda_1 [(01)(12)\cdots(d-2\; d-1) ]$.  The construction of
{\tt CINC} then takes place in the following steps:
\begin{itemize}
\item  We may write $\mbox{exp}(-iH_{int} \pi/(\hbar\Omega)) = 
\Lambda_1(I_{d-2} \oplus \sigma^z)$.
\item  We next argue that we may construct
$\Lambda_1[(j \; j+1)]$.  Indeed, first note that using an
appropriate single qudit permutation matrix $U_{j,k}$, we may construct
$\Lambda_1( I_{j} \oplus \sigma^z \oplus I_{d-2-j})$ as
\begin{equation}
\begin{array}{lll}
\quad\quad \Lambda_1( I_{j} \oplus \sigma^z \oplus I_{d-2-j}) &=& 
I_d \otimes U_{j+2,d-1}(\pi/2,0)\\
& & \Lambda_1(I_{d-2} \oplus \sigma^z) \\
& &I_d \otimes U_{j+2,d-1}(-\pi/2,0).\;
\end{array}
\end{equation}
Then
\begin{equation}
\begin{array}{lll}
\quad\quad\Lambda_1[(j \; j+1)]&=& I_d\otimes U_{j+1,j+2}(\pi/4,\pi/2) \\
& &
\Lambda_1( I_{j} \oplus \sigma^z \oplus I_{d-2-j})\\
& & I_d\otimes U_{j+1,j+2}(-\pi/4,\pi/2).
\end{array}
\end{equation}
\item This leads to the realization of {\tt CINC} in a maximum of $d-1$ controlled
operations,
given that
\hbox{${\tt CINC} \; = 
\; \Lambda_1(01) \Lambda_1(12) \cdots \Lambda_1(d-2 \;d-1)$.}
\end{itemize}

We finally consider the construction of an arbitrary $\Lambda_1(v)$ for
$v \in U(d)$.  Again using standard Givens arguments, it suffices to
construct $\Lambda_1 [ I_j \oplus w \oplus I_{d-j-2}]$ for any
$w \in U(2)$, $\mbox{det}(w)=1$.  Indeed, using the block-wise permutation
argument above, $\Lambda_1( w \oplus I_{d-2})$ suffices.
Now recall (\cite{BarencoEtAl:95}, Lemma 5.1) that there exists
for any $w$ as above factor matrices $a$, $b$, and $c$ such that
$w = a\sigma^x b \sigma^x c$ while $I_2 = abc$.  Hence
\begin{equation}
\Lambda_1(w \oplus I_{d-2}) \ = \ 
(I_d \otimes a) \Lambda_1[(01)] (I_d \otimes b) \Lambda_1[(01)]
(I_d \otimes c)
\end{equation}
This completes the bootstrap argument for exactly universaility, when
the restricted one-qudit Hamiltonian set $\{H_{jk}^{x,y}\}$ is augmented
by $H_{int}$.

We showed how the gate {\tt CINC} can be constructed using the entangling interaction $H_{int}$.
In many situations, the interaction between qudits
will contain more than one term on the diagonal.  For instance, the true Hamiltonian 
may be 
\begin{equation}
H_{int}^{\prime}=\sum_{mn=0}^{d-1} \hbar\Omega_{mn} \ket{mn}\bra{mn}.\;
\end{equation}
In this case the interaction is entangling iff the 
following is true \cite{Brylinski:02} 
\begin{equation}
\Omega_{mn}+\Omega_{pq}\neq \Omega_{mq}+\Omega_{pn} 
+ 2\pi k \quad \mbox{ for any } m,n,p,q, \mbox{ any }k \in \mathbb{Z}.\;
\end{equation}
When the interaction $H_{int}^{\prime}$ is entangling, it is always possible to map it to $H_{int}$ using multiple applications of the coupling conjugated by single qudit gates.  In practice, some multiqudit operations may be done more efficiently using $H^{\prime}_{int}$ directly.

There are several proposals for realizing 
diagonal coupling gates in real physical systems.  For example, in trapped atoms
possible coupling mechanisms include pairwise interactions via dipole-dipole interactions 
\cite{Brennen:QC1,Jaksch:QgateRydberg}, and controlled ground state-ground state 
collisions \cite{Jaksch:Qgateoptlatt}.  The later proposal has been realized recently between atoms 
trapped in an optical lattice \cite{Greiner}.  These proposals were originally made with 
the goal of engineering two qubit controlled phase gates.  As such, a na\"ive adaptation to
encoding over all magnetic hyperfine levels would fail due to off diagonal couplings between 
basis states.  However, it 
should be possible to modify one or more proposals to realize a differential shift on a 
single product state.  For instance, in Ref. \cite{Stock:Coll} it was proposed to realize
a quantum gate using the ground state-ground state collisional shift induced by shape 
resonance.  Here one can tune a magnetic field such that a single molecular 
state is on resonance with a bound motional state of an external trap for both atoms.  Because the 
resonance is dependent on the internal states, a unique 
phase is accumulated on a single product state.  Provided 
the atoms are sufficiently separated, the other basis state pairs do not 
interact and a Hamiltonian of the form $H_{int}$ is realized (up to local unitaries.)  

\section{Conclusions}
\label{sec:conc}
We have identified the criteria for exact quantum computation in qudits.  Our method is constructive and relies on the $QR$ decomposition of unitaries on qudits using a gate library generated by a fixed set of single qudit Hamiltonians and one two qudit entangling gate.  Using the concept of a coupling graph we are able to show that universal computation is possible if the nodes (equivalently logical basis states) are connected.  Further we give a prescription for efficient single qudit computation by demanding that at each stage of the $QR$ decomposition the graph remain connected.  Using the gate library generated by the couplings in Eq. \ref{singlecouple} the maximum number of gates is $3d(d+1)/2-3$.  The technique for computation is exemplified with a quoctet using the Raman coupled magnetic sublevels of $^{87}$Rb.  It is shown that arbitary single quoctet computation is possible with at most $k=49$ laser pulse sequences.  This gate count is optimal and could be reduced to the minimum number $k_{min}=d(d-1)/2=28$ only if one appends the diagonal generators $H_{jk}^z$ to the library of couping Hamiltonians.      

We note that while the results herein have focused on the construction of unitaries, the ideas can be extended to simulating non-unitary processes such as generalized measurements.  Generalized measurements on a system $s$ can be
thought of as orthogonal measurements on an extended system
$H_{s}\oplus H_{s}^{\perp}$, which may not be orthogonal in 
$s$ alone.  Applications including
precision measurement \cite{Helstrom:StateEst}, quantum communication in the 
context of entanglement
purification \cite{Bennett:Entpur}, and quantum error correction 
\cite{Preskill:QECreq}.  To realize this positive operator valued measurement (POVM),
one can perform a unitary operation on $H_{s}\oplus H_{s}^{\perp}$ followed by 
a projective measurement on $H_{s}^{\perp}$ alone.  For example, non-orthogonal measurements on a qubit can be realized by appending ancillary qubits, performing unitary operations on the joint system, and measuring the ancillae.  The requirement of using two qubit gates can be obviated if the ancillary degrees of freedom come from orthogonal states within the same system.  For example, one can use the $d-2$ states of a qudit to implement POVMs on an orthogonal qubit subspace.  These ideas are explored in the context of quantum optical systems in  Refs. \cite{Arnold:01,Brennen:01}.  The techniques reported here indicate that the requisite operations on the appended Hilbert space can be done efficiently.  

\acknowledgements
GKB appreciates helpful discussions with Ivan Deutsch.  This work is supported in part by a grant from ARDA/NSA.

\appendix
\section{Connection to earlier multiqudit gate constructions}
\label{sec:conn}

Mathukrishnan and Stroud have shown \cite{MathukrishnanStroud:00} that exact universal computation
over qudits can be achieved using a gate library containing a $2d-1$ parameter
family of two qudit gates.  We show that this family of gates can be simulated using 
$d-1$ instances of a single parameter two-qudit gate generated by the Hamiltonian $H_{int}$ (Eq. \ref{couple}).

They begin by writing the unitary $W\in SU(d^n)$ in its spectral decomposition:
\begin{equation}
W=\sum_{j=0}^{d^n-1}e^{i\lambda_j}|\lambda_j\rangle\langle \lambda_j|.
\end{equation}  
The unitary can then be expressed as the product 
\begin{equation}
W=\prod_{j=0}^{d^n-1}X_f(|\lambda_j\rangle)
V_f(\lambda_j)X_f^{-1}(|\lambda_j\rangle),\;
\end{equation}
Here the operator $V_f(\lambda_j)$ applies a phase 
only to a fiducial logical basis state $\ket{f}$,
\begin{equation}
V_f(\lambda_j)=e^{i\lambda_j}|f\rangle\langle f|
+\sum_{k\neq f}|k\rangle\langle k|,\;
\label{statephase}
\end{equation}
The operator $X_f(|\lambda_j\rangle)$ is a unitary extension of the map from the fiducial state to an eigenvector of $U$:
\begin{equation}
X_j(|\lambda_j\rangle)=|\lambda_j\rangle\langle f|
+\sum_{k\neq f}|\chi_k(j)\rangle\langle k|.\;
\label{staterot}
\end{equation}
where $\langle \lambda_j|\chi_k(j)\rangle=0$ and $\bra{\chi_k(j)}\chi_{k^{\prime}}(j)\rangle=\delta_{k,k^{\prime}}$.  There is a freedom in the choice 
of the unitary extension by fixing the set of mappings $\{\ket{k}\rightarrow\ket{\chi_k(j)}\}$.  Notice that arbitrary single qudit operations can be constructed using the spectral decomposition for $n=1$ and choosing the fiducial state to be logical basis state of one qudit.  Here we fix $\ket{f}=\ket{d-1}$.  The two multiqudit operators Eqs. \ref{statephase},\ \ref{staterot} can be simulated exactly the using single qudit operations and two families of controlled two-qudit operators.  The first two-qudit gate defines a one parameter family of 
controlled-phase gates and is in fact generated directly by $H_{int}$:
\begin{equation}
\Lambda_{1}(V(\phi))=e^{-iH_{int}\phi/(\hbar\Omega)}.\;
\label{cphase}
\end{equation}
The second family of operators is defined
$\Lambda_{1}(X(|\psi\rangle))$ and maps $|d-1\rangle\rightarrow|\psi\rangle$ 
on the target qudit iff the control is in state $|d-1\rangle$ and applies 
${\bf 1}$ to the target otherwise:
\begin{equation}
\begin{array}{lll}
\Lambda_{1}(X(|\psi\rangle))&=&\sum_{k\neq d-1,k^{\prime}}
|kk^{\prime}\rangle\langle kk^{\prime}|\\
& &+|d-1\rangle\langle d-1|\otimes (|\psi\rangle\langle d-1|\\
& &+\sum_{k\neq d-1}|\beta_k\rangle\langle k|),\;
\end{array}
\end{equation}
where $\langle\psi|\beta_k\rangle=0$ and 
$\langle\beta_j|\beta_k\rangle=\delta_{j,k}$.  Because the gate is allowed to 
implement any unitary extension of $|\psi\rangle_t\langle d-1|$, it only 
depends on the $2d-2$ parameters of the state $|\psi\rangle$ (two parameters 
are fixed by the norm $\langle\psi|\psi\rangle=1$ and setting the global 
phase to zero.)  

The gate $\Lambda_{1}(X(|\psi\rangle))$ can be simulated exactly with the 
controlled-phase gate (Eq. \ref{cphase}) and single qudit gates as we now show.  First, expand the state
$\ket{\psi}$ in the single qudit basis:  $|\psi\rangle=\sum_{j=0}^{d-1}c_j|j\rangle$, where the global phase is chosen so that $\arg{c_{d-1}}=1$.  The 
conditional mapping $|d-1\rangle\rightarrow|\psi\rangle$ can be realized as a 
sequence of $d-1$ controlled unitaries that couple two target qudit basis 
states at a time,
\begin{equation}
\Lambda_{1}(X(|\psi\rangle))=\prod_{j=0}^{d-2}
\Lambda_1(U_{j,d-1}(\gamma_j,\phi_j)),\;
\end{equation}

The arguments $(\theta_j,\phi_j)$ for each controlled unitary must satisfy 
the following relations:
\begin{equation}
\begin{array}{lll}
c_{d-2}&\equiv&\langle d-2|U_{d-2,d-1}|d-1\rangle=
-ie^{i\phi_{d-2}}\sin{\gamma_{d-2}}\\
c_{d-3}&\equiv&\langle d-3|U_{d-3,d-1}U_{d-2,d-1}|d-1\rangle\\
&=&\langle d-3|U_{d-3,d-1}|d-1\rangle\\
& &\langle d-1|U_{d-2,d-1}|d-1\rangle\\
&=&-ie^{-\phi_{d-3}}\sin{\gamma_{d-3}}\cos{\gamma_{d-2}}\\
&\vdots&\\
c_{k}&=&-ie^{i\phi_k}\sin{\gamma_k}\prod_{l=k+1}^{d-2}
\cos{\gamma_{l}}\ \ \ (k<d-2).\;
\end{array}
\end{equation}
Now it only remains to demonstrate that each controlled rotation 
$\Lambda_1(U_{j,d-1})$ can be simulated with just the controlled phase 
gate and rotations on the target qudit.  A single conjugation suffices:
\begin{equation}
\begin{array}{lll}
\Lambda_1(U_{j,d-1})&=&{\bf 1}\otimes U_{j,d-1}(\gamma_j/2+\pi,\phi_j)\\
& &e^{-iH_{int}\pi/(\hbar\Omega)}{\bf 1}\otimes U_{j,d-1}(-\gamma_j/2+\pi,\phi_j).\;
\end{array}
\end{equation}
Following this construction, $d-1$ controlled phase gates and $d$ single 
qudit gates suffice to exactly simulate $\Lambda_1(X(|\psi\rangle))$.


\begin{thebibliography}{99}

\bibitem{MathukrishnanStroud:00}
A. Mathukrishnan and C.R.Stroud Jr.,
Phys. Rev. A {\bf 62}, 052309 (2000).

\bibitem{Blume-KahoutEtAl:02}
R. Blume-Kahout, C.M. Caves, I.H. Deutsch, 
Found. Phys. {\bf 32}, 1641 (2002).

\bibitem{Aharonov}
D.~Aharonov, Presented at {\em Conference on Quantum Information:  Entanglement, Decoherence and Chaos},   Institute for Theoretical Physics, Santa Barbara, 2001 (unpublished).

\bibitem{Brylinski:02}
J.-L. Brylinski and R.~Brylinski, Mathematics of Quantum Computation, edited by R.~Brylinski and G.~Chen, CRC Press (2002). quant-ph/0108062.

\bibitem{NielsenChuang:00}
      M. Nielsen and I. Chuang,
      {\em Quantum Computation and Quantum Information},
      Cambridge Univ. Press, 2000.

\bibitem{Cybenko:01}
G. Cybenko, 
Reducing quantum computations to elementary unitary operations,
Comp. in Sci. and Eng., 27, March/April 2001.

\bibitem{BullockMarkov:04}  
S.S.~Bullock and I.L.~Markov, Quant. Info. and Comp. {\bf 4}, 27 (2004).

\bibitem{footnote}
The equality of the Lande-g factors up to a sign is an approximation that neglects the nuclear magneton.  For $^{87}$Rb this approximation is good to within $0.1\%$ \cite{Steck} but for larger nuclei such as $^{133}$Cs the error is non-negligible.  The correction does not affect the results here.

\bibitem{Steck}
D.A.~Steck, {\em Rubidium 87 D Line Data}, document available online at {\tt http://steck.us/alkalidata}.

\bibitem{O'Leary}
D.P.~O'Leary and S.S.~Bullock, submitted to
{\em Electronic Transactions in Numerical Analysis},\\
{\tt http://math.nist.gov/$\tilde{\;}$SBullock}.

\bibitem{BarencoEtAl:95}
A. Barenco et al.,
Elementary gates for quantum computation,
Phys. Rev A, {\bf 52} 3457 (1995).

\bibitem{Grassl:03}
M.~Grassl, M.~Roetteler, and T.~Beth, Int. J. Found. of Comp. Sci., {\bf 14}, 757 (2003).
 
\bibitem{Brennen:QC1}
G.K.~Brennen, I.H.~Deutsch,and C.J.~Williams, Phys. Rev. A {\bf 65}, 022313 (2002). 
 
 \bibitem{Jaksch:QgateRydberg}
 D.~Jaksch, J.I.~Cirac, P.~Zoller, S.L.~Rolston, R.~Cote, and M.D.~Lukin, Phys. Rev. Lett. {\bf 85}, 2208 (2000).
 
 \bibitem{Jaksch:Qgateoptlatt}
 D. Jaksch, H.J.~Briegel, J.I.~Cirac, C.W.~Gardiner, and P.~Zoller, Phys. Rev. Lett. {\bf 82} 1975 (1999). 
  
\bibitem{Greiner}
A.~Widera, O.~Mandel, M.~Greiner, S.~Kreim, T.W.~Hansch, and I.~Bloch, Phys. Rev. Lett. {\bf 92}, 160406 (2004).
  
 \bibitem{Stock:Coll}
 R.~Stock, E.L..~Bolda, and I.H.~Deutsch, Phys. Rev. Lett. {\bf 91}, 183201 (2003).
 
  \bibitem{Helstrom:StateEst}
 C.W.~Helstrom, {\em Quantum Detection and Estimation Theory} (Academic Press, New York, 1976).
 
 \bibitem{Bennett:Entpur}
 C.H.~Bennett, D.P.~DiVincenzo, J.A.~Smolin, and W.K.~Wootters, Phys. Rev. A {\bf 54}, 3824 (1996).
 
 \bibitem{Preskill:QECreq}
 J.~Preskill, Proc. R. Soc. London Ser. A {\bf 454}, 385 (1998).
 
  \bibitem{Arnold:01}
 S.~Franke-Arnold, E.~Andersson, S.M.~Barnett, and S.~Stenholm, Phys. Rev. A {\bf 63}, 052301 (2001).
 
 \bibitem{Brennen:01}
 G.K.~Brennen, Ph.D. Thesis, University of New Mexico (2001).
 
\end{thebibliography}
\end{document}